\documentclass{article}
\usepackage{graphics}
\usepackage{graphicx}
\usepackage{amssymb}
\usepackage{amsmath}
\usepackage{amsfonts}
\usepackage{parskip}
\usepackage{cite}
\usepackage{epsfig}
\usepackage{float}
\usepackage{fontenc}
\usepackage[latin1]{inputenc}
\usepackage{times}
\usepackage[english]{babel}

\begin{document}
\title{Non-isothermal transport of multi-phase fluids in porous media. \\
The entropy production}
\author{Signe Kjelstrup$^a$, Dick Bedeaux$^a$, Alex Hansen$^b$, Bj{\o}rn
Hafskjold$^a$, Olav Galteland$^a$ \\
PoreLab, $^a$Department of Chemistry, $^b$Department of Physics \\
Norwegian University of Science and Technology, NTNU, 7491 Trondheim}
\maketitle

\begin{abstract}
We derive the entropy production for transport of multi-phase fluids in a
non-deformable, porous medium exposed to differences in pressure,
temperature, and chemical potentials. Thermodynamic extensive variables on
the macro-scale are obtained by integrating over a representative elementary
volume (REV). Using Euler homogeneity of the first order, we obtain the
Gibbs equation for the REV. From this we define the intensive variables, the
temperature, pressure and chemical potentials and, using the balance
equations, derive the entropy production for the REV. The entropy production
defines sets of independent conjugate thermodynamic fluxes and forces in the
standard way. The transport of two-phase flow of immiscible components is
used to illustrate the equations.
\end{abstract}



\noindent Key words: porous media, energy dissipation, two-phase flow,
excess surface- and line-energies, pore-scale, representative elementary
volume, macro-scale, non-equilibrium thermodynamics

\section{Introduction}


The aim of this article is to develop a macro-scale description of
multi-phase flow in porous media in terms of non-equilibrium thermodynamics.
The system consists of several fluid phases in a medium of constant
porosity. The aim is to describe transport on the scale of measurements; 
\emph{i.e.} on the macro-scale, using properties defined on this scale,
which represent the underlying structure on the micro-scale. The effort is
not new; it was pioneered more than 30 years ago \cite%
{Bedford1983,hg1990,gh1998,Hilfer1998}, and we shall build heavily on these
results, in particular those of Hassanizadeh and Gray \cite{hg1990,gh1998}.
The aim is also still the original one; to obtain a systematic description,
which can avoid arbitrariness and capture the essential properties of
multi-component multi-phase flow-systems. Not only bulk energies need be
taken into account to achieve this. Also the excess surface- and
line-energies must be considered.

But, unlike what has been done before, we shall seek to reduce drastically
the number of variables needed for the description, allowing us still to
make use of the systematic theory of non-equilibrium thermodynamics. While
the entropy production in the porous medium so far has been written as a
combination of contributions from each phase, interface and contact line, we
shall write the property for a more limited set of macro-scale variables.
This will enable us to describe experiments and connect variables at this
scale.

The theory of non-equilibrium thermodynamics was set up by Onsager \cite%
{Onsager1931a, Onsager1931b} and further developed for \emph{homogeneous}
systems during the middle of the last century \cite{deGroot1984}. It was the
favored thermodynamic basis of Hassanizadeh and Gray for their description
of porous media. These authors \cite{hg1990,gh1998} discussed also other
approaches, \emph{e.g} the theory of mixtures in macroscopic continuum
mechanics, cf. \cite{Bedford1983,Hilfer1998}.

The theory of classical non-equilibrium thermodynamics has been extended to
deal with a particular case of flow in heterogeneous systems, namely
transport along \cite{Bedeaux1976} and perpendicular \cite{Kjelstrup2008} to
layered interfaces. A description of heterogeneous systems \emph{on the
macro-scale} has not been given, however. Transport in porous media take
place, not only under pressure gradients. Temperature gradients will
frequently follow from transport of mass, for instance in heterogeneous
catalysis \cite{Zhu2006}, in polymer electrolyte fuel cells, in batteries 
\cite{Kjelstrup2008,Richter2017}, or in capillaries in frozen soils during
frost heave \cite{Forland1981}. The number of this type of phenomena is
enormous. We have chosen to consider only the vectorial driving forces
related to changes in pressure, chemical composition and temperature,
staying away for the time being from deformations, chemical reactions, or
forces leading to stress \cite{Huyghe2017}. The multi-phase flow problem is
thus in focus. 

The development of a general thermodynamic basis for multi-phase flow \cite%
{hg1990,gh1998} started by introduction of thermodynamic properties for each
component in each phase, interface and three-phase contact line. A
representative volume element (REV) was introduced, consisting of bulk
phases, interfaces and three-phase contact lines. Balance equations were
formulated for each phase in the REV, and the total REV entropy production
was the sum of the separate contributions from each phase.

In a recent work by Hansen et al. \cite{Hansen2018}, it was recognized that
the description of the motion of the fluids at the coarse-grained level
could be described by extensive variables. The properties of Euler
homogeneous functions could then be used to create relations between the
flow rates at this level of description. This work, however, did not address
the coarse-graining problem itself. We shall take advantage of Euler
homogeneity also here and use it in the coarse-graining process described
above.

Like Hassanizadeh and Gray \cite{hg1990,gh1998} we use the entropy
production as the governing property. But rather than dealing with the total
entropy production as a sum of numerous parts, we shall seek to define the
total entropy production directly from a basis set of a few coarse-grained
variables. Central in the work is the REV, essentially the same as defined
before. With the REV as a starting point we will derive the entropy
production in this part of the work. The independent and measurable
variables are defined here. The independent constitutive equations that
follow from the entropy production will be given in a next part of the work. Doing this, we
aim to contribute towards solving the scaling problem; i.e. how a
macro-level description can be obtained consistent with the micro-level one.
The overall aim is to describe transport due to thermal, chemical,
mechanical and gravitational forces in a systematic, course-grained manner
that is sufficiently simple for practical use.

\section{System}

\begin{figure}[tbp]
\centering
\includegraphics[scale= 0.35]{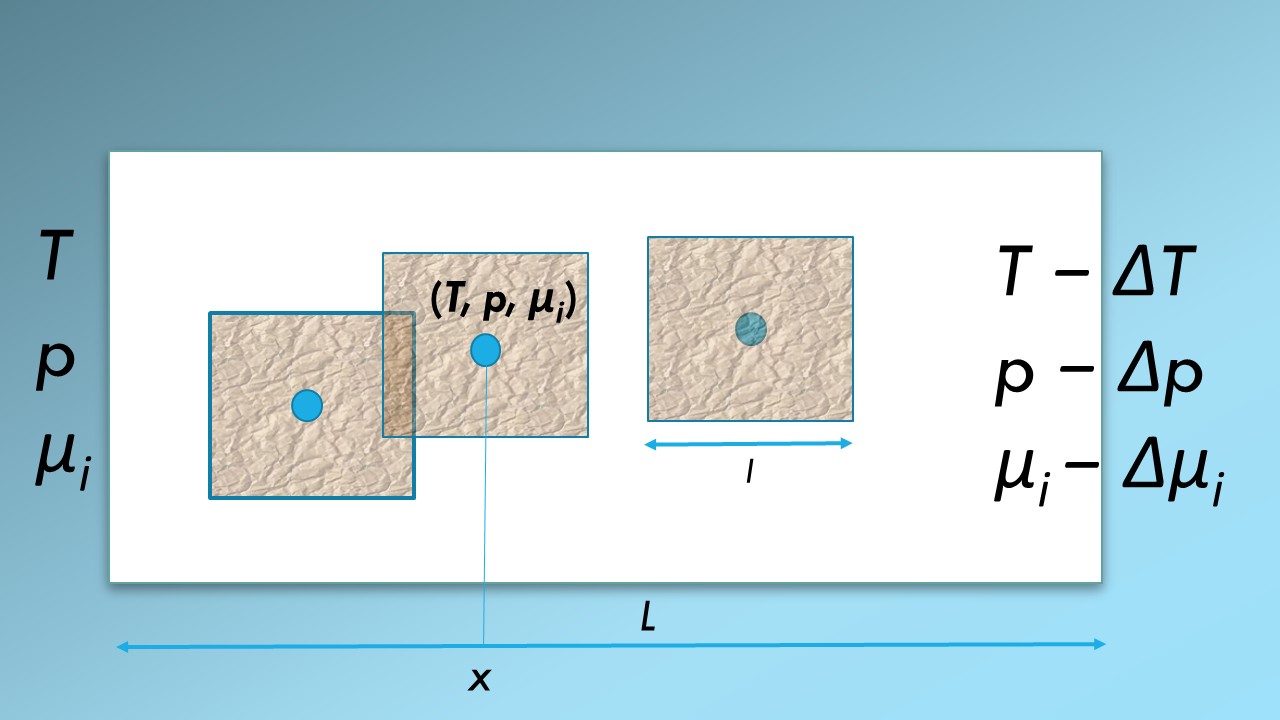}
\caption{Schematic illustration of a heterogeneous system (white box, length 
$L$) exposed to a difference in temperature, $\Delta T$, pressure, $\Delta p$
or chemical potential $\Delta \protect\mu_i$. The system is isolated in the $%
y,z-$directions. Net flows take place in the $x-$direction. Three
representative elementary volumes, REVs (magenta squares, length $l$) are
indicated. The REVs may overlap. Each is represented by a set of variables $%
(p,T,\protect\mu_i)$ which defines a state (blue dot). Such states can be
defined anywhere on the $x$-axis.}
\label{fig:REV2}
\end{figure}

\begin{figure}[tbp]
\centering
\includegraphics[scale = 0.30]{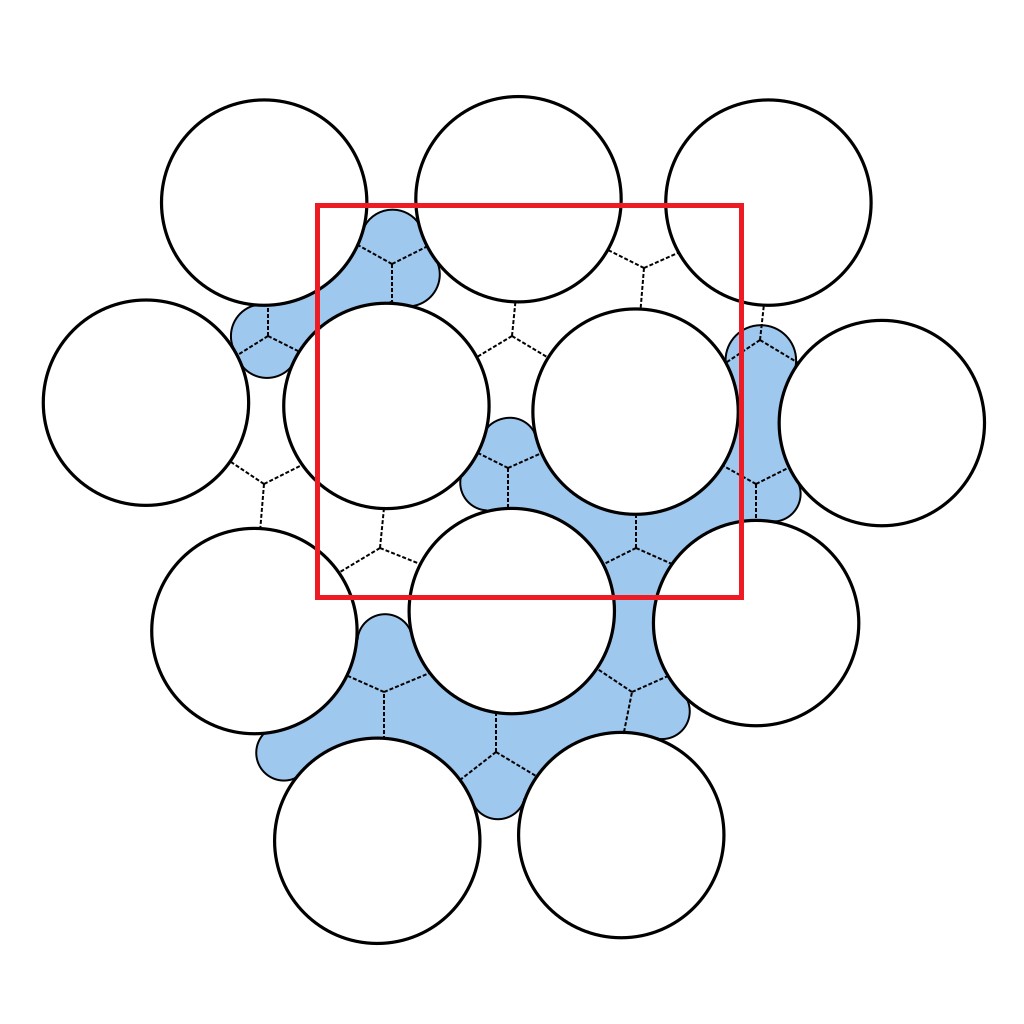}
\caption{A collection of pores filled with two immiscible liquids. In order
to compute the system properties, we define a REV. A REV is indicated in the
figure by the red square. Courtesy of M. Vassvik}
\label{fig:network_REV}
\end{figure}

Consider a heterogeneous system as illustrated by the (white) box in Fig. %
\ref{fig:REV2}. 
The system is a porous medium of fixed porosity filled with several
immiscible fluids. There is net transport in one direction only, the $x$%
-direction. On the scale of measurement, the system is without structure. By
zooming in, we see the pore scale. A collection of pores with two fluids is
schematically shown in Figure \ref{fig:network_REV}. 

A temperature, pressure or chemical potential difference is applied between
the inlet and the outlet, and these differences can be measured. The
pressure difference $\Delta p$ between the outlet and the inlet was defined
for steady state conditions by Tallakstad et al. \cite{Tallakstad2009}, as
the time average of the fluctuating difference $\Delta p(t)$: 
\begin{equation}
\Delta p=\frac{1}{t_{\mathrm{e}}-t_{\mathrm{b}}}\int_{t_{\mathrm{b}}}^{t_{%
\mathrm{e}}}\Delta p(t)dt.
\end{equation}
Here $t$ is the time. Subscript 'b' denotes beginning and 'e' denotes the
end of the measurement. We adopt similar definitions for $\Delta T$ and $%
\Delta \mu _{i}$. It is possible, through application of separate inlet
channels, to control the flow into and out of the system and find the flow
of each component, to define the flow situation in Fig. \ref{fig:REV2}. In
the presence of two immiscible phases, it is only possible to define the
pressure difference between the inlet and the outlet for the phases, $\Delta
p^{w}$ and $\Delta p^{n}$, if there is continuity in the respective phases.

We will repeatedly use two-phase flow as example, where $w$ indicates the
most wetting and $n$ the least wetting phase. We refer to them simply as the
wetting and the non-wetting phase. In most of the paper we consider a
multi-phase fluid. In the system pictured in Fig. \ref{fig:network_REV},
there is flow within the REV in the direction of the pore, which we will
also call a link. This is not necessarily the direction given by the overall
pressure gradient. The flow on the macro-scale, however, is always in the
direction of the pressure gradient. Net flow in other directions are zero
due to isolation of the system in these directions. By flow on the
macro-scale, we mean flow in the direction of the overall pressure gradient
along the $x$-coordinate in Fig.\ref{fig:REV2}. The value of this average
flow is of interest.

The representative volume element, REV, is constructed from a collection of
pores like those contained in the red square in Fig.\ref{fig:network_REV}.
In Fig.\ref{fig:REV2}, three REVs are indicated (magenta structured
squares). In a homogeneous system, statistical mechanical distributions of
molecular properties lead to the macroscopic properties of a volume element.
In a heterogeneous system like here, the statistical distributions are over
the states within the REV. The collection of pores in the REV, cf. Fig. \ref%
{fig:network_REV}, should be of a size that is large enough to provide
meaningful values for the extensive variables, and therefore well defined
intensive variables (see below, Eqs. \ref{H.2} and \ref{H.3}), cf. section
3.2 below. Thermodynamic relations can be written for each REV.

State variables characterize the REV. They are represented by the (blue)
dots in Fig. \ref{fig:REV2}. The size of the REV depends on its composition
and other conditions. Typically, the extension of a REV, $l$, is large
compared to the pore size of the medium, and small compared to the full
system length $L$. This construction of a REV is similar to the procedure
followed in smoothed particle hydrodynamics \cite{Monaghan1992}, cf. the
discussion at the end of the work.

The REVs so constructed, can be used to make a path of states, over which we
can integrate across the system. Each REV in the series of states, is
characterized by variables $T,p,\mu _{i}$, as indicated by the blue dots in
Fig.\ref{fig:REV2}. Vice versa, each point in a porous medium can be seen as
a center in a REV. The states are difficult to access directly, but can be
accessed via systems in equilibrium with the states, as is normal in
thermodynamics. This is discussed at the end of the work. We proceed to
define the REV-variables.

\section{Properties of the REV}

\subsection{Porosity and saturation}

\label{sec:geom}

Consider a solid matrix of constant porosity $\phi$. We are dealing with a
class of systems that are homogeneous in the sense that the typical pore
diameter and pore surface area are, on the average, the same everywhere. The
pores are filled with a mixture of $m-1$ fluid phases; the solid matrix is
labeled $m$. Properties will depend on the time, but this will not be
indicated explicitly in the equations.

The system is filled with $m$ phases. In a simple case, the phases are
immiscible, and chemical constituents are synonymous with a phase. The state
of the REV can be characterized by the volumes of the fluid phases $%
V^{\alpha ,\mathrm{REV}}$, $\alpha =1,..,m-1$ and of the solid medium $V^{m,%
\mathrm{REV}}$. The total volume of the pores is 
\begin{equation}
V^{p,\text{REV}}\equiv \sum_{\alpha =1}^{m-1}V^{\alpha ,\text{REV}}.
\label{I.1}
\end{equation}%
while the volume of the REV is 
\begin{equation}
V^{\text{REV}}\equiv V^{m,\mathrm{REV}}+V^{p,\text{REV}}+\sum_{\alpha >\beta
>\delta =1}^{m}V^{\alpha \beta \delta ,\text{REV}}.
\end{equation}%
Superscript REV is used to indicate a property of the REV. The last term is
the sum of the excess volumes of the three-phase contact lines. While the
excess volume of the surfaces is zero by definition, this is not the case
for the three-phase contact lines. The reason is that the dividing surfaces
may cross each other at three lines which have a slightly different
location. The corresponding excess volume is in general very small, and will
from now on be neglected. This gives the simpler expression 
\begin{equation}
V^{\text{REV}}\equiv V^{m,\mathrm{REV}}+V^{p,\text{REV}}.  \label{I.2}
\end{equation}%
All these volumes can be measured.

The porosity, $\phi $, and the saturation, $\hat{S}$, are given by 
\begin{equation}
\phi \equiv \frac{V^{p,\text{REV}}}{V^{\text{REV}}}\ \ \ \ \ \mathrm{{and}\
\ }\hat{S}^{\alpha }\equiv \frac{V^{\alpha ,\text{REV}}}{V^{p,\text{REV}}}=%
\frac{V^{\alpha ,\text{REV}}}{\phi V^{\text{REV}}}.  \label{I.3}
\end{equation}%
The porosity and the saturation are intensive variables. They do not depend
on the size of the REV. They have therefore no superscript. It follows from
these definitions that 
\begin{equation}
\sum_{\alpha =1}^{m-1}\hat{S}^{\alpha }=1\ \ \ \ \ \mathrm{{and}\ \ }V^{m,%
\text{REV}}=\left( 1-\phi \right) V^{\text{REV}}  \label{I.4}
\end{equation}%
In addition to the volumes of the different bulk phases (they are fluids or
solids) $m\geq \alpha \geq 1$, there are interfacial areas, $\Omega $,
between each two phases in the REV: $\Omega ^{\alpha \beta \text{,REV}},\
m\geq \alpha >\beta \geq 1$. The total surface area of the pores is
measurable. It can be split between various contributions 
\begin{equation}
\Omega ^{p\text{,REV}}=\sum_{\alpha =1}^{m-1}\Omega ^{m\alpha \text{,REV}}
\label{I.5}
\end{equation}

When the surface is not completely wetted, we can estimate the surface area
between the solid $m$ and the fluid phase $\alpha $, from the total pore
area available and the saturation of the component. 
\begin{equation}
\Omega ^{m\alpha \text{,REV}}=\hat{S}^{\alpha }\Omega ^{p\text{,REV}}
\label{I.6}
\end{equation}%
This estimate is not correct for strongly wetting components or dispersions.
In those cases, films can be formed at the walls, and $\Omega ^{m\alpha 
\text{,REV}}$ is not proportional to $\hat{S}^{\alpha }$. In the class of
systems we consider, all fluids touch the wall, and there are no films of
one fluid between the wall and another fluid.

\subsection{Thermodynamic properties of the REV}

We proceed to define the thermodynamic properties of the REV within the
volume $V^{\mathrm{REV}}$ described above. In addition to the volume, there
are other additive variables. They are the masses, the energy and the
entropy. We label the components (the chemical constituents) using italic
subscripts. There are in total $n$\ components distributed over the phases,
surfaces and contact lines. The mass of component $i$, $M_{i}^{\text{REV}}$,
in the REV is the sum of bulk masses, $M_{i}^{\alpha ,\text{REV}}$, $m\geq
\alpha \geq 1$, the excess interfacial masses, $M_{i}^{\alpha \beta ,\text{%
REV}}$, $m\geq \alpha >\beta \geq 1$, and the excess line masses, $%
M_{i}^{\alpha \beta \delta ,\text{REV}}$, $m\geq \alpha >\beta >\delta \geq
1 $. 
\begin{equation}
M_{i}^{\text{REV}}=\sum_{\alpha =1}^{m}M_{i}^{\alpha ,\text{REV}%
}+\sum_{\alpha >\beta =1}^{m}M_{i}^{\alpha \beta ,\text{REV}}+\sum_{\alpha
>\beta >\delta =1}^{m}M_{i}^{\alpha \beta \delta ,\text{REV}}  \label{V.1}
\end{equation}%
There is some freedom in how we allocate the mass to the various phases and
interfaces \cite{Gibbs1961,Kjelstrup2008}. We are \textit{e.g.} free to
choose a dividing surface such that one $M_{i}^{\alpha \beta ,\text{REV}}$
equals zero. A zero excess mass will simplify the description, but will
introduce a reference. The dividing surface with zero $M_{i}^{\alpha \beta ,%
\text{REV}}$ is the equimolar surface of component $i$. The total mass of a
component in the REV is, however, \emph{independent} of the location of the
dividing surfaces. From the masses, we compute the various mass densities 
\begin{eqnarray}
\rho _{i} &\equiv &\frac{M_{i}^{\text{REV}}}{V^{\text{REV}}},\ \ \rho
_{i}^{\alpha }\equiv \frac{M_{i}^{\alpha ,\text{REV}}}{V^{\alpha ,\text{REV}}%
},\ \   \notag \\
\rho _{i}^{\alpha \beta } &\equiv &\frac{M_{i}^{\alpha \beta ,\text{REV}}}{%
\Omega ^{\alpha \beta ,\text{REV}}},\ \ \rho _{i}^{\alpha \beta \delta
}\equiv \frac{M_{i}^{\alpha \beta \delta ,\text{REV}}}{\Lambda ^{\alpha
\beta \delta ,\text{REV}}}  \label{V.2}
\end{eqnarray}%
where $\rho _{i}$ and $\rho _{i}^{\alpha }$ have dimension kg.m$^{-3}$, $%
\rho _{i}^{\alpha \beta }$ has dimension kg.m$^{-2}$ and $\rho _{i}^{\alpha
\beta \delta }$ has dimension kg.m$^{-1}$.

All densities are for the REV. If we increase the size of the REV, by for
instance doubling its size, $V^{\text{REV}}$, $M_{i}^{\text{REV}}$ and other
extensive variables will all double. They will double, by doubling all
contributions to these quantities. But this is not the case for the density $%
\rho _{i}$ or the other densities. They remain the same, independent of the
size of the REV. This is true also for the densities of the bulk phases,
surfaces and contact lines. Superscript REV is therefore not used for the
densities.

Within one REV there are natural fluctuations in the densities. But the
densities make it possible to give a description on the macro-scale
independent of the precise size of the REV. The densities will thus be used
in our description on the macro-scale. The density $\rho _{i}^{\alpha }$ may
vary somewhat in $V^{\alpha }$. We can then find $M_{i}^{\alpha }$ as the
integral of $\rho _{i}^{\alpha }$ over $V^{\alpha }$. Equation \ref{V.2}
then gives the volume-averaged densities.

The internal energy of the REV, $U^{\text{REV}}$, is the sum of bulk
internal energies, $U^{\alpha ,\text{REV}}$, $m\geq \alpha \geq 1$, the
excess interfacial internal energies, $U^{\alpha \beta ,\text{REV}}$, $m\geq
\alpha >\beta \geq 1$, and the excess line internal energies, $U^{\alpha
\beta \delta ,\text{REV}}$, $m\geq \alpha >\beta >\delta \geq 1$:%
\begin{equation}
U^{\text{REV}}=\sum_{\alpha =1}^{m}U^{\alpha ,\text{REV}}+\sum_{\alpha
>\beta =1}^{m}U^{\alpha \beta ,\text{REV}}+\sum_{\alpha >\beta >\delta
=1}^{m}U^{\alpha \beta \delta ,\text{REV}}  \label{V.3}
\end{equation}%
The summation is taken over all phases, interfaces and contact lines (if
non-negligible). We shall see in a subsequent paper  how these contributions may give
specific contributions to the driving force. The internal energy densities
are defined by 
\begin{eqnarray}
u & \equiv & \frac{U^{\text{REV}}}{V^{\text{REV}}},\ \ u^{\alpha}\equiv 
\frac{U^{\alpha ,\text{REV}}}{V^{\alpha ,\text{REV}}},\ \   \notag \\
u^{\alpha \beta} &\equiv &\frac{U^{\alpha \beta ,\text{REV}}}{\Omega
^{\alpha \beta ,\text{REV}}},\ \ u^{\alpha \beta \delta }\equiv \frac{%
U^{\alpha \beta \delta ,\text{REV}}}{\Lambda ^{\alpha \beta \delta ,\text{REV%
}}}  \label{V.4}
\end{eqnarray}%
Their dimensions are J.m$^{-3}$ ($u ,u^{\alpha}$), J.m$^{-2}$ ($u^{\alpha
\beta}$) and J.m$^{-1}$ ($u^{\alpha \beta \delta}$), respectively.

The entropy in the REV, $S^{\text{REV}}$, is the sum of bulk entropies, $%
S^{\alpha ,\text{REV}}$, $m\geq \alpha \geq 1$, the\ excess entropies, $%
S^{\alpha \beta ,\text{REV}}$, $m\geq \alpha >\beta \geq 1$, and the excess
line entropies, $S^{\alpha \beta \delta ,\text{REV}}$, $m\geq \alpha >\beta
>\delta \geq 1$:%
\begin{equation}
S^{\text{REV}}=\sum_{\alpha =1}^{m}S^{\alpha ,\text{REV}}+\sum_{\alpha
>\beta =1}^{m}S^{\alpha \beta ,\text{REV}}+\sum_{\alpha >\beta >\delta
=1}^{m}S^{\alpha \beta \delta ,\text{REV}}  \label{V.5}
\end{equation}%
The entropy densities are defined by%
\begin{equation}
s\equiv \frac{S^{\text{REV}}}{V^{\text{REV}}},\ \ s^{\alpha }\equiv \frac{%
S^{\alpha ,\text{REV}}}{V^{\alpha ,\text{REV}}},\ \ s^{\alpha \beta }\equiv 
\frac{S^{\alpha \beta ,\text{REV}}}{\Omega ^{\alpha \beta ,\text{REV}}},\ \
s^{\alpha \beta \delta }\equiv \frac{S^{\alpha \beta \delta ,\text{REV}}}{%
\Lambda ^{\alpha \beta \delta ,\text{REV}}}  \label{V.6}
\end{equation}%
and have the dimensions J{.K}$^{-1}$.m$^{-3}$ ($s,s^{\alpha }$), J{.K}$^{-1}$%
.m$^{-2}$ ($s^{\alpha \beta }$) and J{.K}$^{-1}$.m$^{-1}$ ($s^{\alpha \beta
\delta }$), respectively. The entropy will in addition have a
configurational contribution from the mixture of components (phases) when
there is a distribution of states within the heterogeneous mixture. To
explain this in more detail; consider the example of the entropy of a
two-phase fluid in a long, narrow tube of variable diameter. None of the
fluids are forming films between the tube walls and the other fluid, and the
long tube can be regarded as an ensemble of shorter tubes, each containing a
bubble \cite{Sinha2013}. The configurational term can then be given in terms
of the probability to find the position of a bubble at position $x_{b}$.

For the volume, Eqs.\ref{I.1} and \ref{I.2} apply when the contact lines
give a negligible contribution. The dividing surfaces by definition have no
excess volume. For all the other extensive thermodynamic variables, like the
enthalpy, Helmholtz, Gibbs energies and the grand potential, relations
similar to Eq.\ref{V.3} and \ref{V.5} apply.

To summarize this section; we have defined a basis set of variables for a
class of systems, where these variables are additive in the manner shown.
From the set of REV variables we obtain the densities, $u$, $s$ or $\rho
_{i} $ to describe the heterogeneous system on the macro-scale. A series of
the REVs of this type, is needed for integration across the system, see
section \ref{fig:Scatchard_aa}.

The surface areas and the contact line lengths are not \emph{independent}
variables in this representation of the REV. These variables are, however,
included indirectly through the assumption that the basic variables of the
REV are additive. This means that a REV of a double size has double the
energy, entropy, and mass, but also double the surface areas of various
types and double the line lengths.


\subsection{REV size considerations}

\begin{figure}[tbp]
\centering
\includegraphics[width=0.48\textwidth]{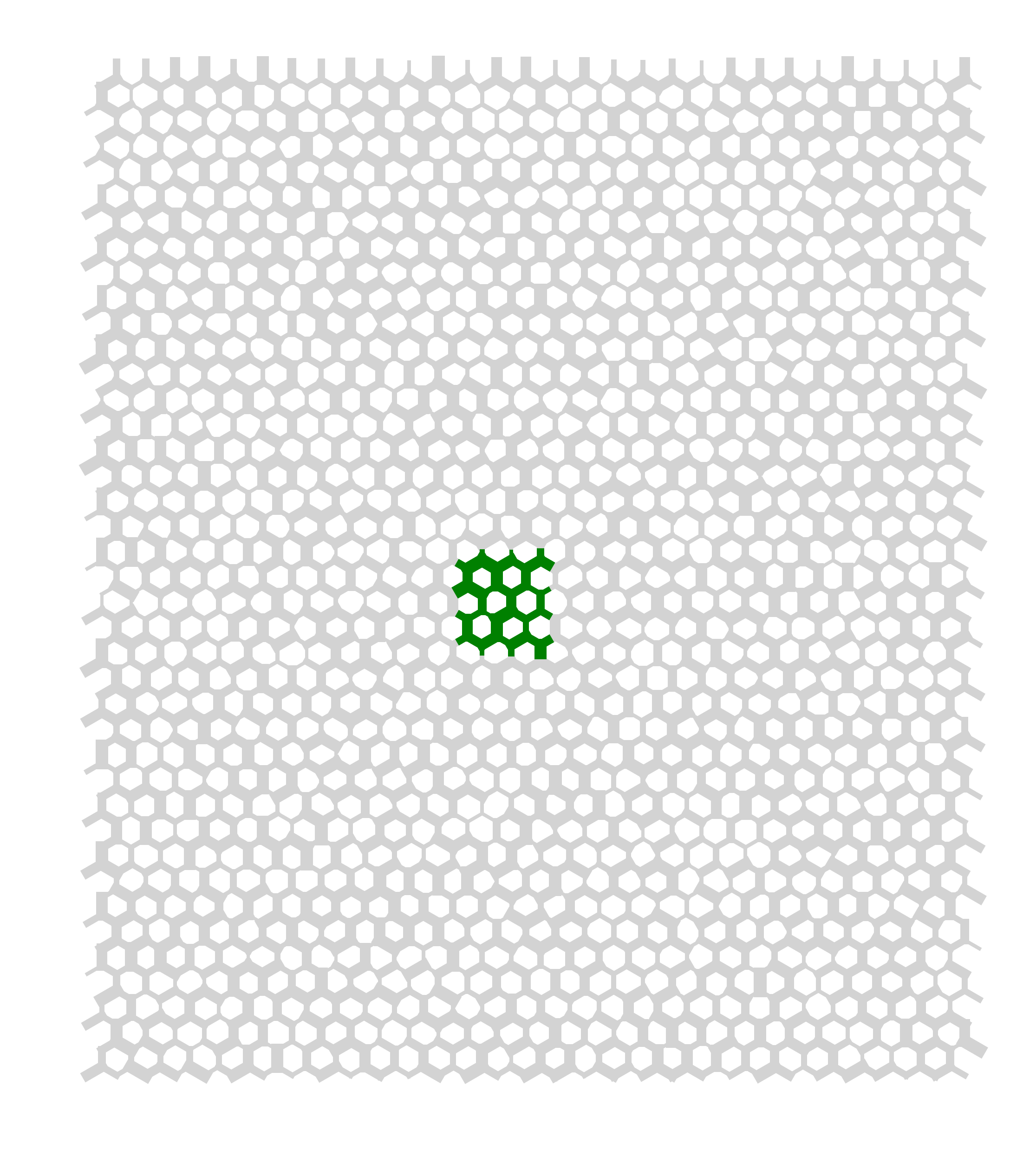} \includegraphics[width=0.48%
\textwidth]{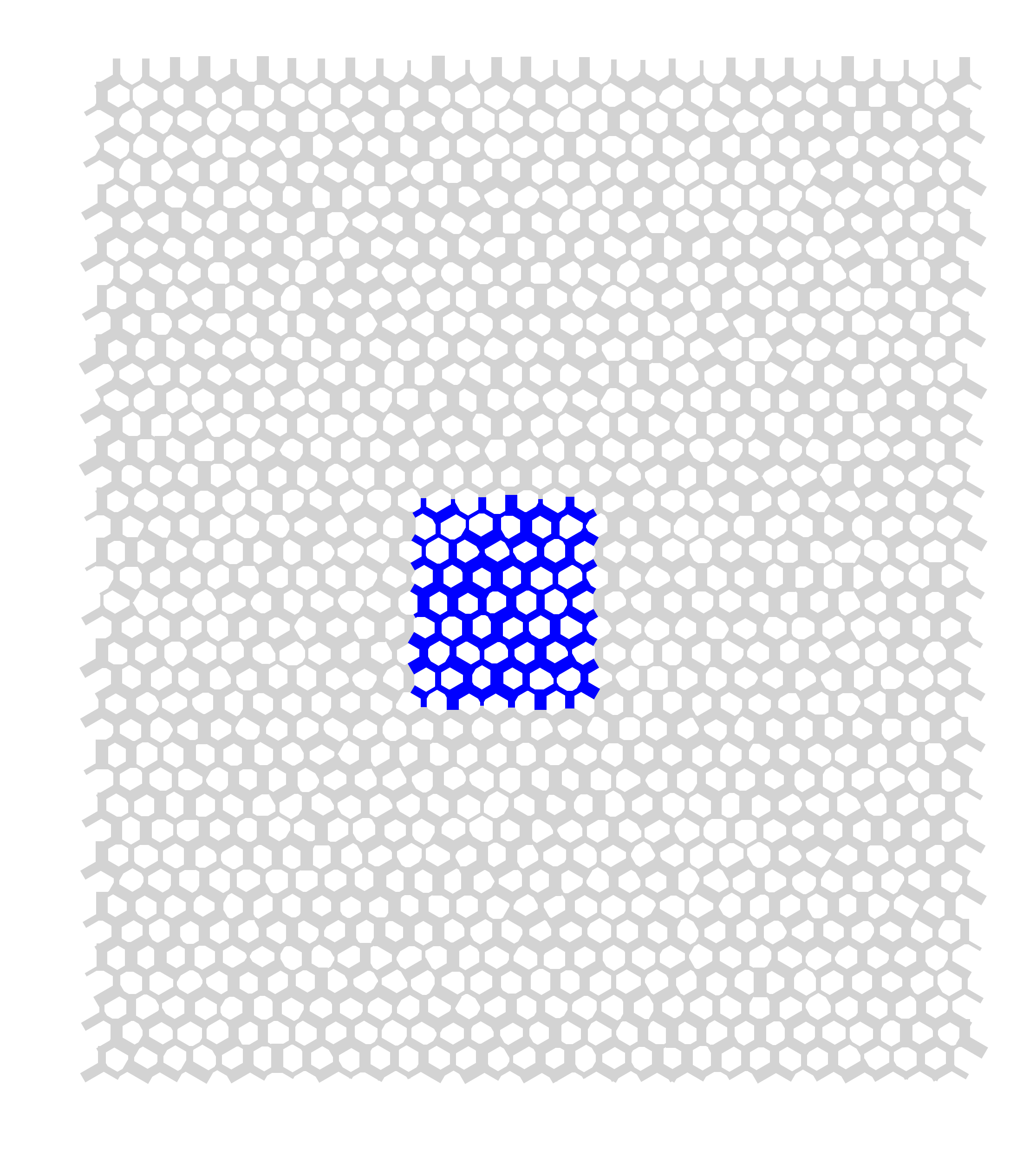}
\caption{Illustration of the link network and two of the candidate REVs
under consideration. The left candidate REV (green) is {5.2}mm $\times$ {6}%
mm and the right candidate REV (blue) is {10.4}mm$\times$ {12}mm.}
\label{fig:network_rev}
\end{figure}

As an illustration of the REV construction, consider the internal energy of
two isothermal, immiscible and incompressible fluids (water and decane)
flowing in a Hele-Shaw cell composed of silicone glass beads. The relevant
properties of the fluids can be found in Table \ref{tab:network_fluids}. The
porous medium is a hexagonal network of 3600 links, as illustrated in Figure %
\ref{fig:network_rev}. The network is periodic in the longitudinal and the
transverse directions and a pressure difference of 1.8 $\times $ 10$^{4}$ Pa
drives the flow in the longitudinal direction. The overall saturation of the
water is 0.4. The network flows were simulated using the method of Aker et
al. \cite{Aker1998}, see Ref.~\cite{Gjennestad2018} for details.

The internal energy of the REV is, according to Section 3.2, a sum over the
two fluid bulk contributions and three interface contributions, 
\begin{align}
U^{\text{REV}}& =U^{m,\text{REV}}+\sum_{i\in \left\{ w,n\right\} }\left\{
U^{i,\text{REV}}\right\} +U^{wn,\text{REV}}+U^{np,\text{REV}}+U^{wp,\text{REV%
}} \\
& =U^{m,\text{REV}}+V^{p,\text{REV}}\sum_{i\in \left\{ w,n\right\} }\left\{ 
\hat{S}^{i}u^{i}\right\} +u^{wn}\Omega ^{wn,\text{REV}}+u^{np}\Omega ^{np,%
\text{REV}}+u^{wp}\Omega ^{wp,\text{REV}}.  \label{eq:network_U_rev}
\end{align}%
where, $u^{i}$ is the internal energy density of phase $i$ and $u^{ij}$ is
the excess internal energy per interfacial area between phase $i$ and phase $%
j$. We assume $u^{i}$\ and $u^{ij}$\ to be constant. For simplicity, $u^{ij}$
is approximated by interfacial tension, denoted $\gamma ^{ij}$. The internal
energy of the porous matrix is constant in this example and is therefore set
to zero. 

\begin{table}[tbp]
\caption{Fluid properties used to compute the candidate REV internal energy,
for a network containing water (n) and decane (w) within silicone glass (p)
at atmospheric pressure and 293 K.}
\label{tab:network_fluids}
\centering
\begin{tabular}{llll}
Parameter & Value & Unit & Reference \\ \hline
$\eta^w$ & 9.2 $\times$ 10$^{-4}$ & Pa.s & \cite{NISTChemistryWebBook} \\ 
$\eta^n$ & 1.0 $\times$ 10$^{-3}$ & Pa.s & \cite{NISTChemistryWebBook} \\ 
$\gamma^{wp}$ & 2.4 $\times$ 10$^{-2}$ & N.m$^{-1}$ & \cite{Neumann1974} \\ 
$\gamma^{np}$ & 7.3 $\times$ 10$^{-2}$ & N.m$^{-1}$ & \cite{Neumann1974} \\ 
$\gamma^{wn}$ & 5.2 $\times$ 10$^{-2}$ & N.m$^{-1}$ & \cite{Zeppieri2001} \\ 
$-u^w$ & 2.8 $\times$ 10$^{8}$ & J.m$^{-3}$ & \cite{NISTChemistryWebBook} \\ 
$-u^n$ & 3.4 $\times$ 10$^{8}$ & J.m$^{-3}$ & \cite{NISTChemistryWebBook} \\ 
\hline
&  &  & 
\end{tabular}%
\end{table}

Candidate REVs are of different sizes, see Table \ref{tab:network_results}.
The 5.4 mm $\times $ 6 mm (green), and 10.4 mm $\times $ 12 mm (blue)
candidate REVs are shown in Figure~\ref{fig:network_rev}. For all candidate
REVs, $U^{\text{REV}}$ is calculated according to Eq.~\ref{eq:network_U_rev}
at each time step. Since the measured saturations and interfacial areas are
fluctuating in time, so is the internal energy. A time-step weighted
histogram of the internal energy presents the probability distribution.

\begin{figure}[tbp]
\centering
\includegraphics[width=0.48\textwidth]{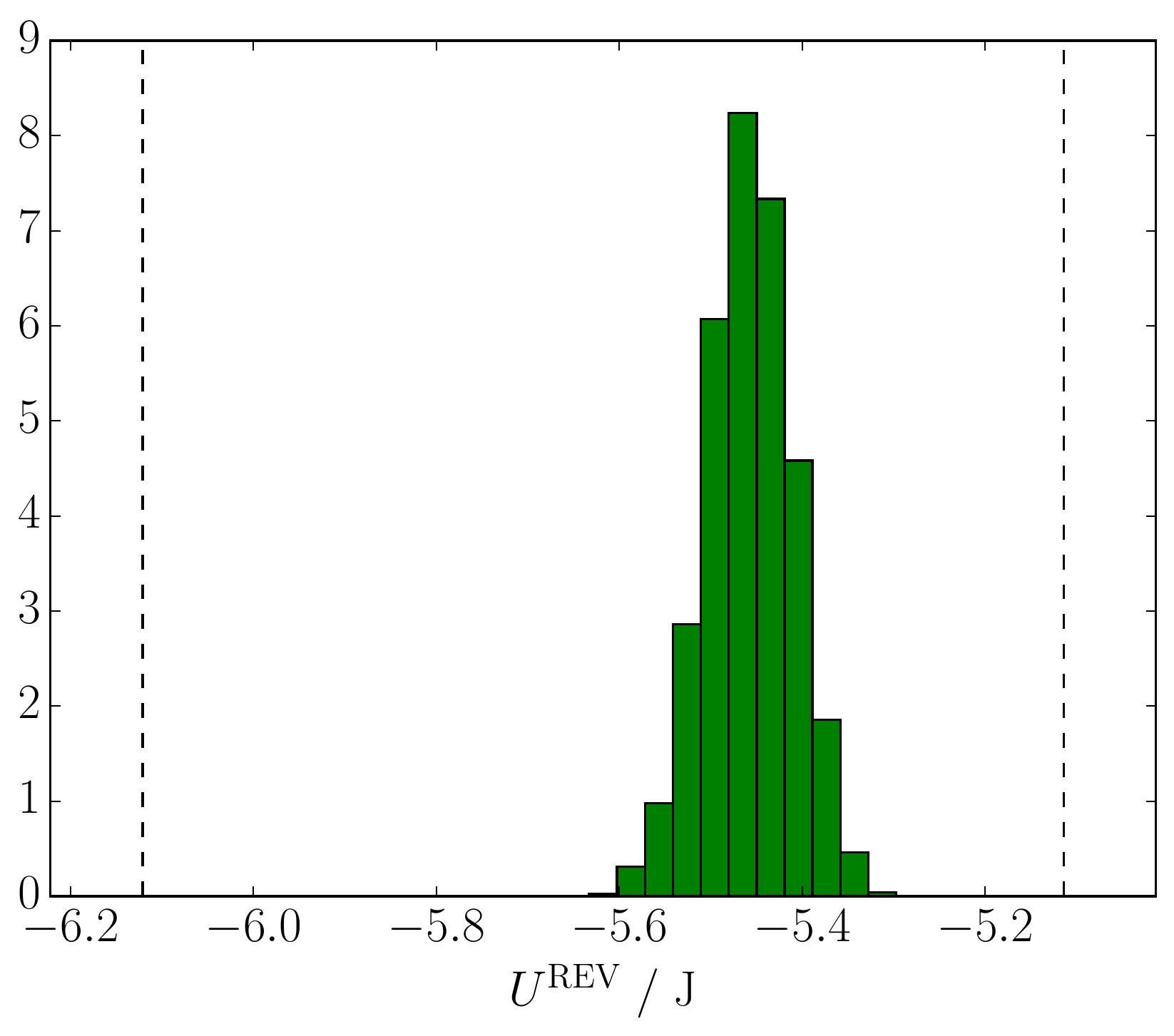} \includegraphics[width=0.47%
\textwidth]{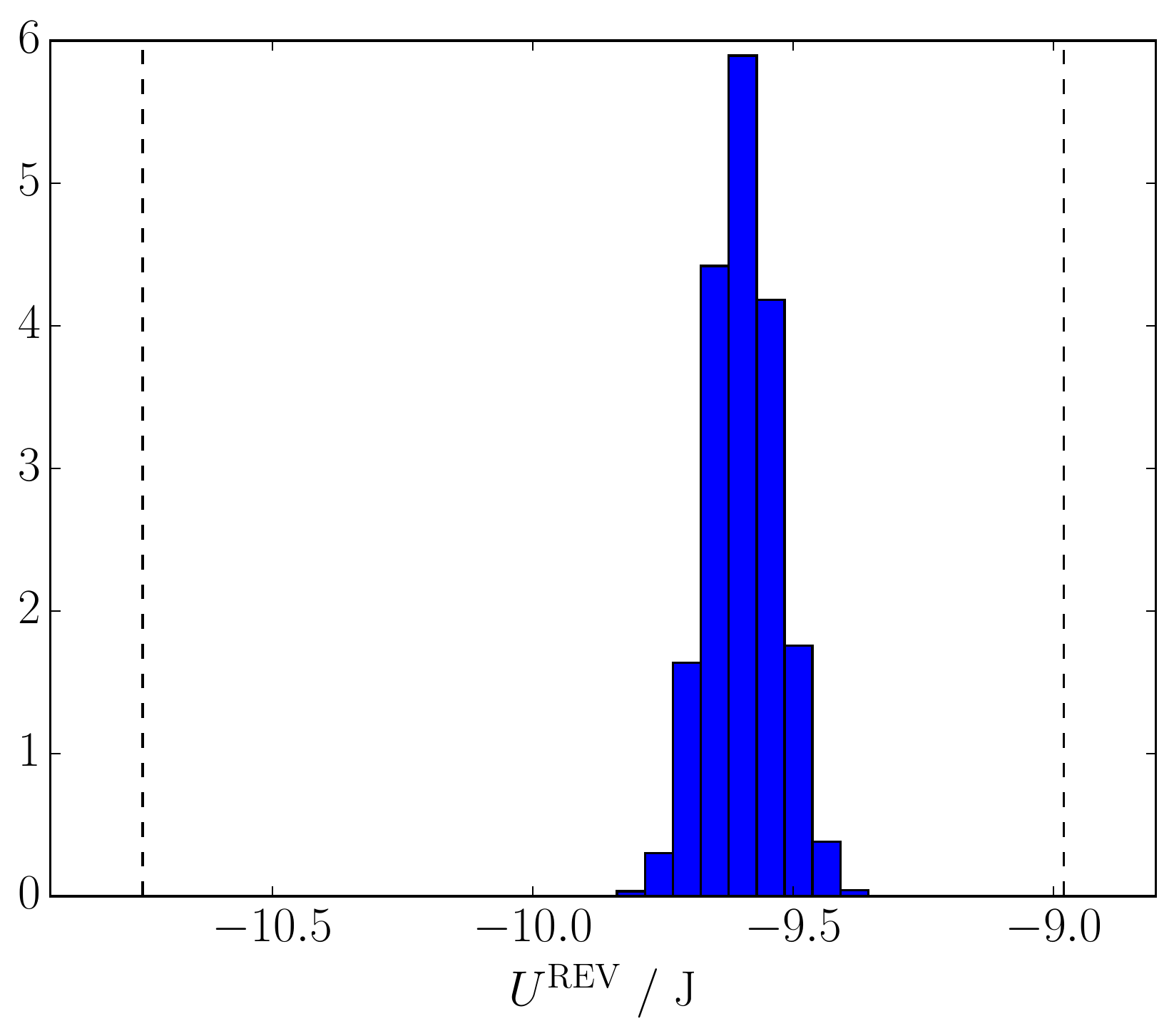}
\caption{Probability distributions of internal energy in two the candidate
REVs. The left candidate REV (green) is {5.2}mm $\times {6}$ mm and the
right candidate REV (blue) is {10.4} mm$\times {12}$ mm. In each plot, the
left dashed line represents the internal energy the candidate REV would have
if it contained non-wetting fluid only and the right dashed line represents
the internal energy the candidate REV would have if it contained wetting
fluid only.}
\label{fig:network_u}
\end{figure}

The probability distributions of $U^{\text{REV}}$ are shown in Figure~\ref%
{fig:network_u} for the 5.2 mm $\times $ 6 mm (green) and 10.4 mm $\times $
12 mm (blue) candidate REVs. In both plots, the vertical lines represent the
internal energy the REV would have if it were occupied by one of the fluids
alone. We denote the difference in internal energy between these two
single-phase states by $\Delta U^{\text{REV}}$.

\begin{table}[tbp]
\caption{Mean values of $U^{\text{REV}}$ and $u$ for candidate REVs of
different sizes, along with the standard deviation of $U^{\text{REV}}$
divided by $\Delta U^{\text{REV}}$. The latter quantity represents a measure
of the relative size of the fluctuations in $U^{\text{REV}}$.}
\label{tab:network_results}
\centering
\vspace{5 mm} 
\begin{tabular}{llll}
Candidate REV & mean $\left( U^{\text{REV}} \right)$ & $\pm \left( U^{\text{%
REV}} \right)$ & mean $\left( u \right)$ \\ 
Size & /J & /$\Delta U^{\text{REV}}$ & / 10$^7$ J m$^{-3}$ \\ \hline
5.2 mm $\times$ {6.0} mm & {-2.82} & {0.069}{} & {-6.04} \\ 
{7.8} mm $\times$ {9.0} mm & {-5.46} & {0.047}{} & {-5.19} \\ 
{10.4} mm $\times$ {12.0} mm & {-9.60} & {0.037}{} & {-5.13} \\ 
{13.0} mm $\times$ {15.0} mm & {-15.4} & {0.028}{} & {-5.25} \\ 
{15.6} mm $\times$ {18.0} mm & {-22.2} & {0.024}{} & {-5.27} \\ 
{18.3} mm $\times$ {21.0} mm & {-29.9} & {0.021}{} & {-5.22} \\ 
{20.8} mm $\times$ {24.0}mm & {-39.1} & {0.017}{} & {-5.23} \\ \hline
\end{tabular}%
\end{table}

The mean value of the $U^{\text{REV}}$ for all candidate REVs are given in
Table~\ref{tab:network_results}, along with mean density $u=U^{\text{REV}%
}/V^{\text{REV}}$ and the standard deviation of $U^{\text{REV}}$ divided by $%
\Delta U^{\text{REV}}$. The latter quantity is a measure of the relative
size of the fluctuations in $U^{\text{REV}}$. Due to the additivity of $U^{%
\text{REV}}$, the mean values of $U^{\text{REV}}$ increases roughly
proportional to the candidate REV size. But this happens only after the REV
has reached a minimum size, here {7.8} mm $\times$ {9.0} mm. For the larger
candidate REVs, the mean value of $u$ changes little as the size increases.
The relative size of the fluctuations in $U^{\text{REV}}$ decreases in
proportion to the linear size of the candidate REVs.

This example indicates that it makes sense to characterize the internal
energy of a porous medium in terms of an internal energy density as defined
by Eq.~\ref{V.3}, given that the size of the REV is appropriately large.
About 100 links seem to be enough in this case. This will vary with the type
of porous medium, cf. the 2D square network model of Savani et al. \cite%
{Savani2017}.


\section{Homogeneity on the macro-scale}

\label{Homogeneity}

Before we address any transport problems, consider again the system pictured
in Fig.\ref{fig:REV2} (the white box). All REVs have variables and densities
as explained above. By integrating to a somewhat larger volume $V$, using
the densities defined, we obtain the set of basis variables, ($U,S,M_{i}$),
in $V$. The internal energy $U$ of the system is an Euler homogeneous
function of first order in $S,V,M_{i}$: 
\begin{equation}
U\left( \lambda S,\lambda V,\lambda M_{i}\right) =\lambda U\left(
S,V,M_{i}\right)  \label{H.1}
\end{equation}%
The internal energy $U,$ volume $V$, entropy $S$, and component mass $M_{i}$%
, obey therefore the Gibbs equation; 
\begin{equation}
dU=\left( \frac{\partial U}{\partial S}\right) _{V,M_{i}}dS+\left( \frac{%
\partial U}{\partial V}\right) _{S,M_{i}}dV+\sum_{i=1}^{n}\left( \frac{%
\partial U}{\partial M_{i}}\right) _{S,V,M_{j}}dM_{i}  \label{H.2}
\end{equation}%
No special notation is used here to indicate that $U,S,V,M_{i}$ are
properties on the macro-scale. Given the heterogeneous nature on the
micro-scale, the internal energy has contributions from all parts of the
volume $V$, including from the excess surface and line energies. By writing
Eq.\ref{H.1} we find that the normal thermodynamic relations apply for the
heterogeneous system at equilibrium, additive properties $U,S,V,M_{i}$,
obtained from sums of the bulk-, excess surface- and excess
line-contributions.

We can then move one more step and use Gibbs equation to \emph{define} the
temperature, the pressure and chemical potentials on the macro-scale as
partial derivatives of $U$: 
\begin{equation}
T\equiv \left( \frac{\partial U}{\partial S}\right) _{V,M_{i}},\ \ p\equiv
-\left( \frac{\partial U}{\partial V}\right) _{S,M_{i}},\ \ \mu _{i}\equiv
\left( \frac{\partial U}{\partial M_{i}}\right) _{S,V,M_{j}}  \label{H.3}
\end{equation}%
The temperature, pressure and chemical potentials on the macro-scale are,
with these formulas, defined as partial derivatives of the internal energy.
This is normal in thermodynamics, but the meaning is now extended. In a
normal homogeneous, isotropic system at equilibrium, the temperature,
pressure and chemical equilibrium refer to a homogeneous volume element. The
temperature of the REV is a temperature representing all phases, interfaces
and lines combined, and the chemical potential of $i$ is similarly obtained
from the internal energy of all phases. Therefore, there are only one $T,p$
and $\mu _{i}$ for the REV. The state can be represented by the (blue) dots
in Fig.1.

On the single pore level, the pressure and temperature in the REV will have
a distribution. In the two immiscible-fluid-example the pressure, for
instance, will vary between a wetting and a non-wetting phase because of the
capillary pressure. One may also envision that small phase changes in one
component (\emph{\emph{e.g.}} water) leads to temperature variations due to
condensation or evaporation. Variations in temperature will follow changes
in composition.

The intensive properties are \emph{not} averages of the corresponding
entities on the pore-scale over the REV. This was pointed out already by
Gray and Hassanizadeh \cite{gh1998}. The definitions are derived from the
total internal energy only, and this makes them uniquely defined. It is
interesting that the intensive variables do not depend on how we split the
energy into into bulk and surface terms inside the REV.

By substituting Eq.\ref{H.3} into Eq.\ref{H.2} we obtain the Gibbs equation
for a change in total internal energy on the macro-scale 
\begin{equation}
dU=TdS-pdV+\sum_{i=1}^{n}\mu _{i}dM_{i}  \label{H.4}
\end{equation}%
As a consequence of the condition of homogeneity of the first order, we also
have 
\begin{equation}
U=TS-pV+\sum_{i=1}^{n}\mu _{i}M_{i}  \label{H.5}
\end{equation}

The partial derivatives $T,\ p$\ and $\mu _{i}$\ are homogeneous functions
of the zeroth order. This implies that%
\begin{equation}
T(\lambda S,\lambda V,\lambda M_{i})=T(S,V,M_{i})  \label{H.6}
\end{equation}%
Choosing $\lambda =1/V$ it follows that 
\begin{equation}
T(S,V,M_{i})=T(s,1,\rho _{i})=T(s,\rho _{i})  \label{H.7}
\end{equation}%
The temperature therefore depends only on the subset of variables $s\equiv
S/V,\rho _{i}\equiv M_{i}/V$ and not on the complete set of variables $%
S,V,M_{i}$. The same is true for the pressure, $p$, and the chemical
potentials, $\mu _{i}$. This implies that $T,\ p$\ and $\mu _{i}$ are not
independent. We proceed to repeat the standard derivation of the Gibbs-Duhem
equation which makes their interdependency explicit.

The Gibbs equation on the macro-scale in terms of the densities follows
using Eqs. \ref{H.4} and \ref{H.5} 
\begin{equation}
du=Tds+\sum_{i=1}^{n}\mu _{i}d\rho _{i}  \label{H.8}
\end{equation}%
which can alternatively be written as%
\begin{equation}
ds=\frac{1}{T}du-\frac{1}{T}\sum_{i=1}^{n}\mu _{i}d\rho _{i}  \label{H.9}
\end{equation}%
The Euler equation implies%
\begin{equation}
u=Ts-p+\sum_{i=1}^{n}\mu _{i}\rho _{i}  \label{H.10}
\end{equation}%
%
%
%
%
%
%
%
%

By differentiating Eq. (\ref{H.10}) and subtracting the Gibbs equation (\ref%
{H.8}), we obtain in the usual way the Gibbs-Duhem equation: 
\begin{equation*}
dp=sdT+\sum_{i=1}^{n}\rho _{i}d\mu _{i}
\end{equation*}%
This equation makes it possible to calculate $p$ as a function of $T$\ and $%
\mu _{i}$ and shows how these quantities depend on one another.

We have now described the heterogeneous porous medium by a limited set of
coarse-grained thermodynamic variables. These average variables and their
corresponding temperature, pressure and chemical potentials, describe a
coarse-grained homogeneous mixture with variables which reflect the
properties of the class of porous media. In standard equilibrium
thermodynamics, Gibbs' equation applies to a homogeneous phase. We have
extended this use to be applicable for heterogeneous systems at the
macro-scale. Whether or not the chosen procedure is viable, remains to be
tested. We refer to the Section 7 of this paper.

\section{Entropy production in porous media}

Gradients in mass- and energy densities produce changes in the variables on
the macro-scale. These lead to transport of heat and mass. Our aim is to
find the equations that govern this transport across the REV. We therefore
expose the system to driving forces and return to Fig. \ref{fig:REV2}.

The balance equations for masses and internal energy of a REV are 
\begin{eqnarray}
\frac{\partial \rho _{i}}{\partial t} &=&-\frac{\partial }{\partial x}J_{i}
\label{E.2} \\
\frac{\partial u}{\partial t} &=&-\frac{\partial }{\partial x}J_{u}=-\frac{%
\partial }{\partial x}\left[ J_{q}^{^{\prime }}+\sum_{i=1}^{n}J_{i}H_{i}%
\right]  \label{E.3}
\end{eqnarray}%
The transport on this scale is in the $x-$direction only. The mass fluxes, $%
J_{i}$, and the flux of internal energy, $J_{u}$, are all macro-scale
fluxes. The internal energy flux is the sum of the measurable (or sensible)
heat flux, $J_{q}^{^{\prime }}$ and the partial specific enthalpy (latent
heat), $H_{i}$ (in J.kg$^{-1}$) times the component fluxes, $J_{i}$, see 
\cite{deGroot1984,gh1998,Kjelstrup2008} for further explanations. Component $%
m$ (the porous medium) is not moving and is the convenient frame of
reference for the fluxes.

The entropy balance on the macro-scale is 
\begin{equation}
\frac{\partial s}{\partial t}=-\frac{\partial }{\partial x}J_{s}+\sigma
\label{E.4}
\end{equation}%
Here $J_{s}$ is the entropy flux, and $\sigma $ is the entropy production
which is positive definite, $\sigma \geq 0$ (the second law of
thermodynamics). We can now derive the expression for $\sigma$ in the
standard way \cite{deGroot1984,Kjelstrup2008}, by combining the balance
equations with Gibbs' equation. The entropy production is the sum of all
contributions within the REV.

In the derivations, we assume that the Gibbs equation is valid for the REV
also when transport takes place. Droplets can form at high flow rates, while
ganglia may occur at low rates. We have seen above that there is a minimum
size of the REV, for which the Gibbs equation can be written. When we assume
that the Gibbs equation applies, we implicitly assume that there exists a
uniquely defined state. The existence of such a state was first postulated
by Hansen and Ramstad \cite{Hansen2009}. Experimental evidence for the
assumption was found by Erpelding \cite{Erpelding2013}.

Under the conditions we demand for the REV, the Gibbs equation Eq.\ref{H.9}
keeps its form during a time interval $dt$, giving 
\begin{equation}
\frac{\partial s}{\partial t}=\frac{1}{T}\frac{\partial u}{\partial t}-\frac{%
1}{T}\sum_{i=1}^{n}\mu _{i}\frac{\partial \rho _{i}}{\partial t}  \label{E.1}
\end{equation}%
We can now introduce the balance equations for mass and energy into this
equation, see \cite{Kjelstrup2008} for details. By comparing the result with
the entropy balance, Eq.\ref{E.4}, we identify first the entropy flux, $%
J_{s} $, 
\begin{equation}
J_{s}=\frac{1}{T}J_{q}^{\prime }+\sum_{i=1}^{n}J_{i}S_{i}  \label{E.5}
\end{equation}%
The entropy flux is composed of the sensible heat flux over the temperature
plus the sum of the specific entropies carried by the components. The form
of the entropy production, $\sigma $, depends on our choice of the energy
flux, $J_{u}$ or $J_{q}^{\prime }$. The choice of form is normally motivated
by practical wishes; what is measurable or computable. We have 
\begin{eqnarray}
\sigma &=&J_{u}\frac{\partial }{\partial x}(\frac{1}{T})-\sum_{i=1}^{n}J_{i}%
\frac{\partial }{\partial x}(\frac{\mu _{i}}{T})  \notag \\
&=&J_{q}^{\prime }\frac{\partial }{\partial x}(\frac{1}{T})-\frac{1}{T}%
\sum_{i=1}^{n}J_{i}\frac{\partial }{\partial x}\mu _{i,T}  \label{E.7}
\end{eqnarray}%
These expressions are equivalent formulations of the same physical
phenomena. When we choose $J_{u}$ as variable with the conjugate force $%
\partial (1/T)/\partial x$, the mass fluxes are driven by minus the gradient
in the Planck potential ${\mu _{i}}/{T}$. When, on the other hand we choose $%
J_{q}^{\prime }$ as a variable with the conjugate force $\partial
(1/T)/\partial x$, the mass fluxes are driven by minus the gradient in the
chemical potential at constant temperature over this temperature. The
entropy production defines the independent thermodynamic driving forces and
their conjugate fluxes. We have given two possible choices above to
demonstrate the flexibility. The last expression is preferred for analysis
of experiments.

In order to find the last line in Eq.\ref{E.7} from the first, we used the
thermodynamic identities $\mu _{i}=H_{i}-TS_{i}$ and $\partial ( \mu
_{i}/T)/\partial (1/T)=H_{i}$ as well as the expression for the energy flux
given in Eq.\ref{E.3}. Here $S_{i}$ is the partial specific entropy (in J.kg$%
^{-1}$.K$^{-1}$).

\subsection{The chemical potential at constant temperature}

The derivative of the chemical potential at constant temperature is needed
in the driving forces in the second line for $\sigma $ in Eq.\ref{E.7}. For
convenience we repeat its relation to the full chemical potential \cite%
{deGroot1984}. The differential of the full chemical potential is: 
\begin{equation}
d\mu _{i}=-S_{i}dT+V_{i}dp+\sum_{j=1}^{n}\left( \frac{\partial \mu _{i}}{%
\partial M_{j}}\right) _{p,T,M_{i}}dM_{j}  \label{eq:mui}
\end{equation}%
where $S_{i},V_{i}$ and $({\partial \mu _{i}}/{\partial M_{j}})_{p,T,M_{i}}$
are partial specific quantities. The partial specific entropy and volume are
equal to: 
\begin{equation}
S_{i}=-\left( \frac{\partial \mu _{i}}{\partial T}\right) _{p,M_{j}}\ ,\
V_{i}=\left( \frac{\partial \mu _{i}}{\partial p}\right) _{T,M_{j}}
\end{equation}%
and the last term of Eq. \ref{E.7} is denoted by 
\begin{equation}
d\mu _{i}^{c}=\sum_{j=1}^{n}\left( \frac{{\partial \mu _{i}}}{{\partial M_{j}%
}}\right) _{p,T,M_{i}}dM_{j}
\end{equation}%
By reshuffling, we have the quantity of interest as the differential of the
full chemical potential plus an entropic term; 
\begin{equation}
d\mu _{i,T}\equiv d\mu _{i}+S_{i}dT=V_{i}dp+d\mu _{i}^{c}  \label{eq:muiT}
\end{equation}
The differential of the chemical potential at constant temperature is 
\begin{equation}
\frac{d\mu _{i,T}}{dx}=\frac{d\mu _{i}^{c}}{dx}+V_{i}\frac{dp}{dx}
\label{eq:muiTb}
\end{equation}

With equilibrium in the gravitational field, the pressure gradient is $%
dp/dx=-\rho g$, where $\rho $ is the total mass density and $g$ is the
acceleration of free fall \cite{Forland1988}. The well known separation of
components in the gravitational field is obtained, with $d\mu _{i,T}=0$ and 
\begin{equation}
\frac{d\mu _{i}^{c}}{dx}=\frac{RT}{W_{i}}\frac{d\ln (\hat{S}_{i}y_{i})}{dx}%
=V_{i}\rho g
\end{equation}%
where $W_{i}$\ is the molar mass (in kg.mol$^{-1}$), $\hat{S}_{i}$ the
saturation, and $y_{i}$ the activity coefficient of component $i$. The gas
constant, $R$, has dimension J.K$^{-1}.$mol$^{-1}$. The gradient of the mole
fraction of methane and decane in the geothermal gradient of the fractured
carbonaceous Ekofisk oil field, was estimated to 5x10$^{-4}$m$^{-1}$ \cite%
{Holt1983}, in qualitative agreement with observations. We replace $d\mu
_{i,T}$ below using these expressions.

It follows from Euler homogeneity that the chemical potentials in a
(quasi-homogeneous) mixture are related by $0=SdT-Vdp+\sum_{j=1}^{n}\rho
_{j}d\mu _{j}$, which is Gibbs-Duhem's equation. By introducing \ref{eq:muiT}
into this equation we obtain an equivalent expression, to be used below: 
\begin{equation}
0=\sum_{j=1}^{n}\rho _{j}d\mu _{j}^{c}  \label{eq:GD}
\end{equation}

\section{Transport of heat and two-phase fluids}

Consider again the case of two immiscible fluids, one more wetting (w) and
one more non-wetting (n). The entropy production in Eq.\ref{E.7} gives, 
\begin{equation}
\sigma =J_{q}^{\prime }\frac{\partial }{\partial x}(\frac{1}{T})-\frac{1}{T}%
\left( J_{w}\frac{\partial \mu _{w,T}}{\partial x}+J_{n}\frac{\partial \mu
_{n,T}}{\partial x}\right)  \label{E.7c}
\end{equation}%
The solid matrix is the frame of reference for transport, $J_{r}=0$ and does
not contribute to the entropy production. The volume flux is frequently
measured, and we wish to introduce this as new variable 
\begin{equation}
J_{V}=J_{n}V_{n}+J_{w}V_{w}
\end{equation}%
Here $J_{V}$ has dimension (m$^{3}$.m$^{-2}.$s$^{-1}$ = m.s$^{-1}$), and the
partial specific volumes have dimension m$^{3}$.kg$^{-1}$. The chemical
potential of the solid matrix may not vary much if the composition of the
solid is constant across the system. We assume that this is the case ($d\mu
_{m}^{c}\approx 0$), and use Eq.\ref{eq:GD} to obtain 
\begin{equation}
0=\rho _{n}d\mu _{n}^{c}+\rho _{w}d\mu _{w}^{c}  \label{GD}
\end{equation}%
The entropy production is invariant to the choice of variables. We can
introduce the relations above and the explicit expression for $d\mu _{i,T}$
into Eq.\ref{E.7c}, and find the practical expression: 
\begin{equation}
\sigma =J_{q}^{\prime }\frac{\partial }{\partial x}\left( \frac{1}{T}\right)
-J_{V}\frac{1}{T}\frac{\partial p}{\partial x}-J_{D}\frac{\rho _{w}}{T}\frac{%
\partial \mu _{w}^{c}}{\partial x}  \label{D.1}
\end{equation}%
In the last line, the difference velocity $J_{D}$ is 
\begin{equation}
J_{D}=\frac{J_{w}}{\rho _{w}}-\frac{J_{n}}{\rho _{n}}
\end{equation}%
This velocity describes the relative movement of the two components within
the porous matrix. In other words, it describes the ability of the medium to
separate components. The main driving force for separation is the chemical
driving force, related to the gradient of the saturation. The equation
implies that also temperature and pressure gradients may play a role for the
separation.

The entropy production has again three terms, one for each independent
driving force. With a single fluid, the number of terms are two. The force
conjugate to the heat flux is again the gradient of the inverse temperature.
The entropy production, in the form we can obtain, Eqs.\ref{E.7c} or \ref%
{D.1}, dictates the constitutive equations of the system.

The volume flows used by Hansen et al \cite{Hansen2018} are related to ours
by $J_{n}V_{n}=Q_{n}/A^{p}$ and $J_{w}V_{w}=Q_{w}/A^{p}$, where $A^{p}\equiv
V^{p}/\ell $ is the cross-sectional area of the pores in the volume normal
to the flow direction, and $\ell $\ is the length of the volume in the flow
direction. In \cite{Hansen2018} the total volume flow was assumed to be an
Euler homogeneous functions of the first order of the fractional areas, $%
A^{n}\equiv V^{n}/\ell $ and $A^{w}\equiv V^{w}/\ell $ for the two
components, respectively. The seepage velocities are $v_{n}=Q_{n}/A^{n}$ and 
$v_{w}=Q_{w}/A^{w}$. By introducing these variables in the expression for
the entropy production, we obtain two alternative forms for the entropy
production, Eq. \ref{D.1}: 
\begin{eqnarray}
&&\sigma =J_{q}^{\prime }\frac{\partial }{\partial x}\left( \frac{1}{T}%
\right) -\frac{1}{A^{p}T}\left[ (Q_{w}+Q_{n})\frac{\partial p}{\partial x}+(%
\frac{Q_{w}}{\rho _{w}V_{w}}-\frac{Q_{n}}{\rho _{n}V_{n}})\rho _{w}\frac{%
\partial \mu _{w}^{c}}{\partial x}\right]  \\
&=&J_{q}^{\prime }\frac{\partial }{\partial x}\left( \frac{1}{T}\right) -%
\frac{1}{T}\left[ v_{w}\hat{S}_{w}\left( \frac{\partial p}{\partial x}+\frac{%
1}{V_{w}}\frac{\partial \mu _{w}^{c}}{\partial x}\right) +v_{n}\hat{S}%
_{n}\left( \frac{\partial p}{\partial x}+\frac{1}{V_{n}}\frac{\partial \mu
_{n}^{c}}{\partial x}\right) \right] 
\end{eqnarray}%
In the second identity we used Gibbs-Duhem, Eq. \ref{GD}. If we define%
\begin{equation}
\frac{\partial p_{i}}{\partial x}\equiv \hat{S}_{i}\left( \frac{\partial p}{%
\partial x}+\frac{1}{V_{i}}\frac{\partial \mu _{i}^{c}}{\partial x}\right) 
\text{ \ \ for \ }i=n,w
\end{equation}%
the entropy production can be written as%
\begin{equation}
\sigma =J_{q}^{\prime }\frac{\partial }{\partial x}\left( \frac{1}{T}\right)
-\frac{1}{T}v_{w}\frac{\partial p_{w}}{\partial x}-\frac{1}{T}v_{n}\frac{%
\partial p_{n}}{\partial x}
\end{equation}%
Whether $p_{w}$\ and $p_{n}$\ have any concrete meaning is not clear. The
difference of their gradients is%
\begin{eqnarray}
\frac{\partial \left( p_{n}-p_{w}\right) }{\partial x} &=&\frac{\hat{S}_{n}}{%
V_{n}}\frac{\partial \mu _{n}^{c}}{\partial x}-\frac{\hat{S}_{w}}{V_{w}}%
\frac{\partial \mu _{w}^{c}}{\partial x}  \notag \\
&=&\left( \frac{\hat{S}_{n}}{V_{n}\rho _{n}}+\frac{\hat{S}_{w}}{V_{w}\rho
_{w}}\right) \rho _{n}\frac{\partial \mu _{n}^{c}}{\partial x}
\end{eqnarray}%
where we again used Gibbs-Duhem in the last identity. As one often
identifies the left hand side of this equation with the gradient of the
capillary pressure, this would relate this force to the chemical force. We
will come back to this issue later.

\subsection{A path of sister systems}

As pointed out above, through the construction of the REV we were able to
create a continuous path through the system, defined by the thermodynamic
variables of the REVs. The path was illustrated by a sequence of dots in
Fig. \ref{fig:REV2}. Such a path must exist, to make integration possible.
Also continuum mixture theory hypothesizes such a path \cite{Hilfer1998}:
Hilfer introduced a series of \emph{mixture} states, to define an
integration path across the porous system, see \emph{e.g.} \cite{Hilfer1998}.

The path created in Section 2 is sufficient as a path integration across the
medium. The access to and measurement of properties in the REVs is another
issue. It is difficult, if not impossible, to measure in situ as stated
upfront. The measurement probe has a minimum extension (of some mm), and the
measurement will represent an average over the surface of the probe. For a
phase with constant density, the average is well defined and measurable. A
link between the state of the REV and a state where measurements are
possible, is therefore needed. We call the state that provides this link a
sister state.

\begin{figure}[tbp]
\centering
\includegraphics[scale= 0.35]{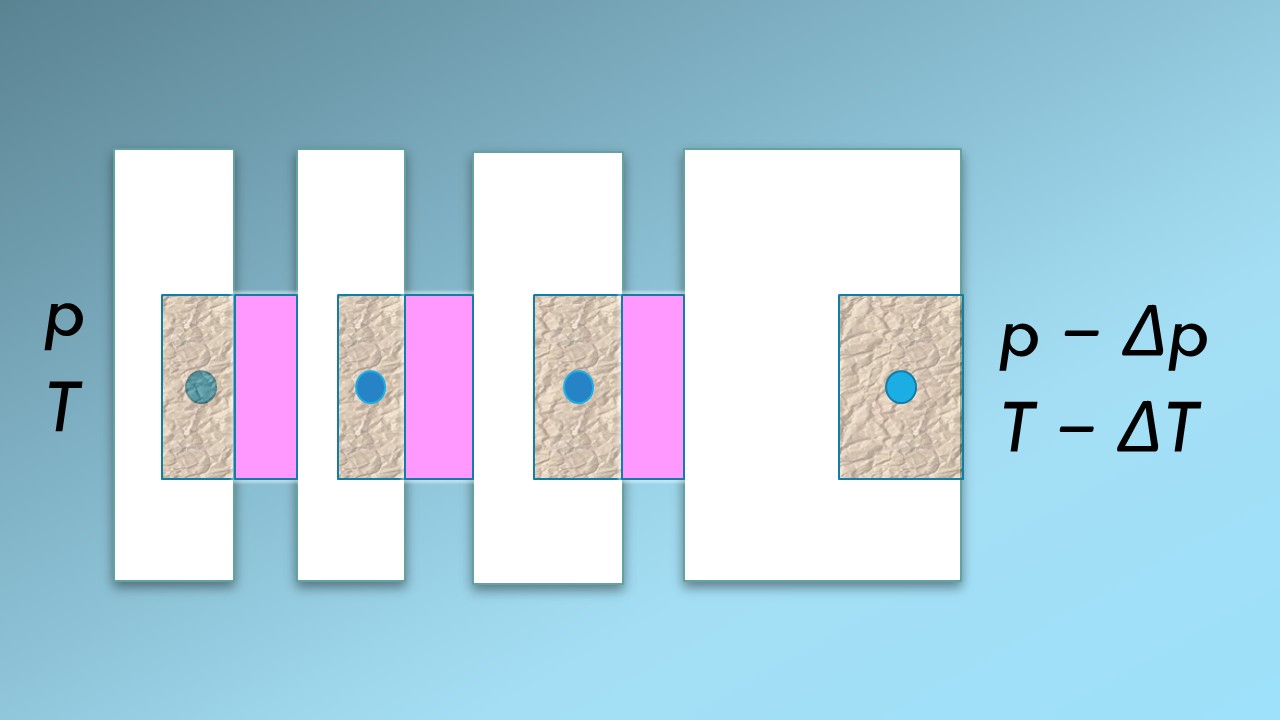}
\caption{A one-dimensional heterogeneous system cut into slices. Each cut is
brought in equilibrium with a homogeneous mixture at the same temperature
and pressure as the REV.}
\label{fig:Scatchard_aa}
\end{figure}

Consider again the path of REVs in the direction of transport. To create the
link between the REV and its sister state, consider the system divided into
slices, see Figure \ref{fig:Scatchard_aa}. The slice (the sister system)
contains homogeneous phases in equilibrium with the REV at the chosen
location.

We hypothesize that we can find such sister states; in the form of a
multi-component mixture with temperature, pressure and composition such that
equilibrium can be obtained with for the REV-variables at any slice
position. The variables of the sister state can then be measured the normal
way. The chemical potential of a component in the sister state can, for
instance, be found by introducing a vapor phase above this state and measure
the partial vapor pressure. We postulate thus that a sister state can be
found, that obey the conditions 
\begin{eqnarray}
T &=&T^{s} \\
p &=&p^{s} \\
\mu _{i} &=&\mu _{i}^{s}  \label{eq:equil}
\end{eqnarray}
Here $i$=1,...n are the components in the REV, and superscript s denotes the
sister state. With the sister states available, we obtain an experimental
handle on the variables of the porous medium. The hypothesis must be
checked, of course.

The series of sister states have the same boundary conditions as the
REV-states, by construction, and the overall driving forces will be the
same. Between the end states, we envision the non-equilibrium system as a
staircase. Each step in the stair made up of REV is in equilibrium with a
step of the sister-state-stair. Unlike the states inside the porous medium,
the sister states are accessible for measurements, or determination of $T,p$
and $\mu_i$. Transport can then be described as taking place along a
sequence of the sister states.

\section{Discussion}


We have seen how we can define a representative elementary volume, called
the REV, and conditions such that standard non-equilibrium thermodynamics
for homogeneous systems \cite{deGroot1984} can be extended to describe flow
in heterogeneous systems like porous media. The procedure adds a step to
earlier elaborations in the field \cite{hg1990,gh1998}. We have been able to
derive the total entropy production directly for the heterogeneous system,
and have thus obtained a new basis for constitutive equations of the system
on the macro-scale. Earlier, the equation sets were written for the separate
parts, or for systems that were homogeneous on the molecular scale \cite%
{Katchalsky1965}.

Equations \ref{E.7c} and \ref{D.1} have been established since long in the
theory of transport in polymer membranes, see \emph{e.g.} \cite%
{Katchalsky1965}. The step to include porous media is inspired by this. The
way of dealing with lack of knowledge of variables inside the system was for
instance used in polymer membrane transport long ago, see \cite%
{Scatchard1953,Laks1964}. The procedure, to introduce a series of
equilibrium states, each state in equilibrium with the membrane at some
location between the external boundaries, was first used by Scatchard \cite%
{Scatchard1953}, and later experimentally verified \cite{Laks1964,Ratkje1988}%
.

The clue to obtain the present results was to use Euler homogeneity on the
macro-scale. This allowed us to set up the Gibbs equation with well defined
variables, in spite of local variations in, for instance, the pressure
inside the REV. Together with the balance equations, this gave the entropy
production in a standard way. Some support can be found in the literature.
Prigogine and Mazur \cite{Prigogine1951} investigated a mixture of two
fluids using non-equilibrium thermodynamics. Their system consisted of
superfluid - and normal helium. Two different pressures were defined. For
the case they discussed, the interaction between the two fluids was small,
meaning that one phase flowed as if the other one (aside from a small
frictional force) was not there. The situation here is similar, as we may
have different liquid pressures inside the REV. But the interaction between
the two immiscible components in our porous medium is large, not negligible
as in the helium case.

A central issue in the extension is the construction of the REV. This
construction is subject to conservation of the important properties of the
system. For instance, the REV must conserve the total porosity of the
system. Such constructions are known from other up-scaling techniques. In
mesoscale simulations, \textit{i.e.} upscaling from properties at the
nanometer scale to micrometer scale, techniques known as "Dissipative
Particle Dynamcs" stemming from hydrodynamics \cite{Hoogerbrugge1992} and
"Smoothed Particle Hydrodynamics", stemming from astrophysics \cite%
{Monaghan1992} have been used.

Inspired by the idea behind smoothed particle hydrodynamics, we can also
define a normalized weight function $W(\mathbf{r})$, such that a microscopic
variable $a(\mathbf{r})$ may be represented by its average, defined as%
\begin{equation}
\overline{a}(\mathbf{r})\equiv \int d\mathbf{r}^{\prime }W(\mathbf{r-r}%
^{\prime })a(\mathbf{r}^{\prime }) .  \label{A.1}
\end{equation}%
For example, if $a(\mathbf{r}^{\prime })$ is the local void fraction in a
porous material as determined from samples of the material, $\overline{a}(%
\mathbf{r})$ is the average porosity of the medium. The average is assigned
to the point $\mathbf{r}$ and varies smoothly in space. The average porosity 
$\overline{a}(\mathbf{r})$ would then be suitable for \textit{e.g.} a
reservoir simulation at the macro-scale.

In general, the system is subject to external forces and its properties are
non-uniform. The choice of $W(\mathbf{r})$ is therefore crucial in that it
defines the extent of the coarse-graining and the profile of the weighting.
The illustration in Fig. 1 alludes to a weight function that is constant
inside a cubic box and zero outside, but other choices are possible. Popular
choices used in mesoscale simulations are the Gaussian and spline functions
(See \cite{Monaghan1992} for details).

A convenient feature of the coarse-graining is that the average of a
gradient of a property $a$ is equal to the gradient of the average. 
\begin{equation}
\nabla \overline{a} (\mathbf{r})=\overline{\nabla a(\mathbf{r})}  \label{A.2}
\end{equation}%
Similarly the average of a divergence of a flux is equal to the divergence
of the average. This implies that balance equations, which usually contain
the divergence of a flux, remain valid after averaging.

Time averages can also be introduced along the lines sketched above. Also
the average of a time derivative of $a$\ becomes equal to the time
derivative of the average. Again, the balance equations remain valid after
averaging. (In this paper we have not used time averages.) The total amount
of $a$\ in the volume of the REV is 
\begin{equation}
A^{\text{REV}}\equiv \int_{\text{REV}}d\mathbf{r}a(\mathbf{r})\equiv 
\overline{a}\int_{\text{REV}}d\mathbf{r}\equiv \overline{a}V^{\text{REV}}
\label{A.3}
\end{equation}

The time scale relevant to porous media transport are usually large
(minutes, hours); and much larger than times relevant for the molecular
scale. Partial derivatives were used in the balance equations and the
entropy production. This signifies that properties can change not only along
the coordinate axis, but also on the time scale. In the present formulation
any change brought about in the REV must retain the validity of the Gibbs
equation (\textit{i.e. }Euler homogeneity). As long as that is true, we may
use the equations, also for transient phenomena.

The outcome of the derivations enables us to deal with a wide range of
non-isothermal phenomena in a systematic manner, from frost heave to
heterogeneous catalysis, or multi-phase flow in porous media. We will
elaborate on what this means in the next part of this work. In particular, we
shall give more details on the meaning of the additive variables and the REV
pressure. We will return to the meaning of the REV variables and how they will
contribute to the driving forces of transport.

\section{Concluding remarks}

We have derived the entropy production for transport of heat and mass in
porous media. The derivations have followed standard non-equilibrium
thermodynamics for heterogeneous systems \cite{Kjelstrup2008}. The only, but
essential, difference has been the fact that we write all these equations
for a porous medium on the macro-scale for the REV of a minimum size using
its total entropy, energy and mass, while these equations often are only
written for the separate contributions. Broadly speaking, we have been
zooming out our view on the porous medium to first define some states that
we take as thermodynamic states because they obey Euler homogeneity. The
states are those illustrated by the dots in Figure \ref{fig:REV2}. In order
to define these states by experiments, we constructed the sister states of
Fig.\ref{fig:Scatchard_aa}.

The advantage of the present formulations is this; it is now possible to
formulate the transport problem on the scale of a flow experiment in
accordance with the second law of thermodynamics, with far less variables.
This also opens up the possibility to test the thermodynamic models for use
along with and compatibility with this law. Such tests can be explicitly
formulated once we have the constitutive equations. In the next part of this work,
such details will be given.

We have been dealing as an example with the specific case of immiscible,
non-isothermal two-phase flow in a non-deformable medium. In deformable
systems, grains may move by translation, compaction and/or rotation. We
expect that this situation also can be treated by the procedure given above.

\subsection*{Acknowledgement}

The authors are grateful to the Research Council of Norway through its
Centres of Excellence funding scheme, project number 262644, PoreLab. Per
Arne Slotte is thanked for stimulating discussions.


{}


\section*{\textbf{Symbol lists}}

\text{\textbf{Table 1. Mathematical symbols, superscripts, subscripts}} 
\newline
\begin{tabular}{ll}
{Symbol} & {Explanation} \\ \hline
$d$ & differential \\ 
$\partial$ & partial derivative \\ 
$\Delta$ & change in a quantity or variable \\ 
$\Sigma$ & sum \\ 
$i$ & subscript meaning component i \\ 
$m$ & number of phases \\ 
$n$ & subscript meaning non-wetting fluid \\ 
$w$ & subscript meaning wetting fluid \\ 
$p$ & superscript meaning pore \\ 
REV & abbreviation meaning representative elementary volume \\ 
$r$ & superscript meaning solid matrix of porous medium \\ 
s & superscript meaning interface \\ 
$u$ & superscripts meaning internal energy \\ 
$\alpha\beta $ & superscripts meaning contact area between phases $%
\alpha\beta $ \\ 
$\alpha\beta\delta$ & superscripts meaning contact line between phases $%
\alpha,\beta,\delta $ \\ 
$\bar{x}$ & average of x \\ \hline
\end{tabular}

\newpage

\textbf{Table 2. Greek symbols, continued} \newline
\begin{tabular}{lll}
{Symbol} & {Dimension} & Explanation \\ \hline
$\alpha$ &  & superscripts meaning a phase \\ 
$\beta$ &  & superscript meaning an interface \\ 
$\delta$ &  & superscript meaning a contact line \\ 
$\phi$ &  & porosity of porous medium \\ 
$\gamma$ & N.m$^{-1}$ (N) & surface tension (line tension) \\ 
$\Lambda$ & m & length of contact line \\ 
$\lambda$ &  & Euler scaling parameter \\ 
$\mu_{\mathrm{i}}$ & J.kg$^{-1}$ & chemical potential of i \\ 
$\rho_{\mathrm{i}} $ & kg.m$^{-3}$ & density, $\equiv M_{\mathrm{i}}/V$ \\ 
$\sigma$ & J.s$^{-1}$.K$^{-1}$.m$^{-3}$ & entropy production in a
homogeneous phase \\ 
$\sigma^{s}$ & J.s$^{-1}$.K$^{-1}$.m$^{-2}$ & surface excess entropy
production \\ 
$\sigma^{c}$ & J.s$^{-1}$.K$^{-1}$.m$^{-1}$ & line excess entropy production
\\ 
$\Omega$ & m$^2$ & surface or interface area \\ \hline
\end{tabular}

\newpage

{\textbf{Table 1. Latin symbols}} \newline
\begin{tabular}{lll}
{Symbol} & {Dimension} & Explanation \\ \hline
$G$ & J & Gibbs energy \\ 
$M$ & kg & mass \\ 
$m$ & kg mol$^{-1}$ &  \\ 
$d$ & m & pore length \\ 
$H_i$ & J kg$^{-1}$ & partial specific enthalpy of $i$ \\ 
$J_i$ & kg.s$^{-1}$.m$^{-2}$ & mass flux of $i$ \\ 
$J_u$ & J.s$^{-1}$.m$^{-2}$ & energy flux \\ 
$J_q^{\prime }$ & J.s$^{-1}$.m$^{-2}$ & sensible heat flux \\ 
$J_V$ & m$^3$.s$^{-1}$.m$^{-2}$ & volume flux \\ 
$l$ & m & characteristic length of representative elementary volume \\ 
$L$ & m & characteristic length of experimental system \\ 
$L_{ij}, \ell_{ij}$ &  & Onsager conductivity \\ 
$p$ & Pa & pressure of REV \\ 
$Q$ & m$^3$.s$^{-1}$ & volume flow \\ 
$\bar{r}$ & m & avarage pore radius \\ 
$S$ & J.K$^{-1}$ & entropy \\ 
$s$ & J.K$^{-1}$.m$^{-3}$ & entropy density \\ 
$S_i$ & J kg$^{-1}$K$^{-1}$ & partial specific entropy of i \\ 
$\hat{S}$ &  & degree of saturation, $\equiv V_{\mathrm{i}}/V$ \\ 
$T$ & K & temperature \\ 
$t$ & s & time \\ 
$U$ & J & internal energy \\ 
$u$ & J.m$^{-3}$ & internal energy density \\ 
$V$ & m$^3$ & volume \\ 
$V_i$ & m$^3$kg$^{-1}$ & partial specific volume \\ 
$x$ & m & coordinate axis \\ 
$x_i$ & - & mass fraction of $i$ \\ 
$W_i$ & - & kg.mol$^{-1}$ molar mass of $i$ \\ \hline
\end{tabular}

\end{document}